# Strategizing COVID-19 Lockdowns Using Mobility Patterns

Olha Buchel[1], Anton Ninkov[2], Danise Cathel[1],
Yaneer Bar-Yam[1], Leila Hedayatifar[1*]

[1]New England Complex Systems Institute, 277 Broadway St, Cambridge, MA, USA
[2]Faculty of Information and Media Studies, University of Western Ontario, Canada

To whom correspondence should be addressed; E-mail: leila@necsi.edu.

**During the COVID-19 pandemic, governments have tried to keep their territories safe by isolating themselves from others, limiting non-essential travel and imposing mandatory quarantines for travelers. While large-scale quarantine has been the most successful short-term policy, it is unsustainable over long periods as it exerts enormous costs on societies. As a result, governments which have been able to partially control the spread of the disease have been deciding to reopen businesses. However, the WHO has warned about the risks of reopening prematurely, as is playing out in some countries such as Spain, France and various states in the US such as California, Florida, Arizona, and Texas. Thus, it is urgent to consider a flexible policy that limits transmission without requiring large scale and damaging quarantines. Here, we have designed a multi-level quarantine process based on the mobility patterns of individuals and the severity of COVID-19 contagion in the US. By identifying the natural boundaries of social mobility, policymakers can impose travel restrictions that are minimally disruptive to social and economic activity. The dynamics of social fragmentation during the COVID-19 outbreak are analyzed by applying the Louvain method with modularity optimization to weekly mobility networks. In a multi-scale community detection process, using the locations of confirmed cases, natural break points as well as high risk areas for contagion are identified. At the smaller scales, for communities with a higher number of confirmed cases, contact tracing and associated quarantine policies is increasingly important and can be informed by the community structure.**

The emergence and global spread of the 2019 novel coronavirus (SARS-CoV-2 or COVID-19) has resulted in a global health emergency. With a high level of observed contagiousness (1) and a lack of proven medical treatment, the situation is becoming increasingly dire as the virus moves across the globe. Public health stakeholders are racing to find adequate methods for intervention as the outbreak spreads (2,3). It is challenging to determine where the next outbreak will be and how to prevent or control it. Analyzing data about positive tests and location of current patients plays a critical role in public health agencies response (4). In most cases, quarantine policies and data related to the COVID-19 outbreak are based on arbitrary borders such as state or county boundary lines (5). While these boundaries may serve constituents well in meeting certain social needs of their communities (e.g. infrastructure, taxes), this is not the most effective way to analyze data for anticipating disease outbreaks.

For the purposes of examining the spread of COVID-19 in the US, mobility patterns can be characterized in three overarching concepts: short distance (e.g., grocery shopping, walking), medium distance (e.g., travel for job or fun), and long distance (e.g., travel to other cities for vacation, visiting families). Travel can be thought of as occurring in "bubbles" of progressively larger geographical scales. National travel bubbles include the common movement of individuals traveling far from one region of the country to another. This type of mobility pattern was quickly identified as risky and attempts to limit it within the country were put in place. For example, in the very beginning of the North American outbreak in March 2020, a group of university students from around the country gathered on Florida beaches for what is known as "Spring Break." During this gathering, local transmission of COVID-19 was detected and the spread of the disease to various other regions of the country occurred (6). Soon after that, most universities were closed and airline travel was reduced. On the other hand, local travel bubbles include regions with close proximity where there are more frequent and consistent mobility patterns. Local bubbles are prevalent in places such as the Northeast Megalopolis (7, 8), where there are numerous cities and communities that all continuously connect to one another. In this region, many individuals live in one city/state (e.g. Philadelphia), work in another (New York City), and vacation in another (New Jersey coast). While these regions are separated by multiple administrative boundaries, they could still be considered to be in the same bubble.

The recent availability of large-scale human activity datasets has greatly improved our ability to study social systems (9–11). Geo-located data sources enable direct observation of social interactions and collective behaviors with unprecedented detail. Networks of human mobility (12–14) have revealed the existence of geo-located communities, or patches that exist at multiple scales from town to city, state, and national scales (15, 16). People in these patches have similar movement patterns and, in a self-organized manner, mostly do not cross the borders of their communities. Borders of patches are subject to vary by changes in the mobility preferences. By spread of coronavirus pandemic in societies, applied quarantine policies and lockdowns on large scales have changed the mobility patterns over the past months. Studying the changes and fragmentation patterns allow us to quantify the effectiveness of the policies and define the risk of the areas based on the mobility of individuals.

In March 2020, technology companies that gather geo-location information on individuals started reporting anonymized mobility data to help researchers stop the spread of the coronavirus. Each data set covers aspects of an individual's movements. Here, we used cell phone data collected by SafeGraph to construct mobility networks from where individuals go. In the mobility network, nodes represent a lattice with cells overlaid on the map of the US. Census block groups are the nodes of the network as the SafeGraph data are aggregated in this level. Edges represent the movement of individuals between two locations (nodes) with a weight represents total number of travels between the locations. We analyze social fragmentation by applying the Louvain method (17) with modularity optimization (18) to the mobility network. Communities refer to the regions in which nodes are more connected to each other than the rest of the network. By adding the number of active confirmed COVID-19 cases to the map, we can define risk exposure for the communities. See supplementary for more detail on data and methods.

After the first cases came to the US through international travels, the spread of COVID-19 has occurred rapidly through patients with or without symptoms at the time of transmission. COVID-19 has an incubation period that typically extends to 14 days with a median time of 4-5 days (19, 20). Movement of asymptomatic individuals in public increases the risk of the disease in the visited areas by them. So, to apply the preventive policies more carefully, it is important to define the geographical patches based on actual mobility of individuals. During the global spread of COVID-19, in the US, state governments are responsible to manage and apply the preventive policies inside their territory. They closed public areas (e.g. work places, universities, schools and shopping centers), and asked people to wear masks. At the beginning of the outbreak, people were also asked to abstain from going outside unless for essential or emergency needs. These actions helped to reduce the unnecessary movements and the spread of the disease. In Figure 1, we compared the fragmentation pattern of the mobility networks of the US for February 23-29 (panel A) and April 5-11 (panel B). Areas with the same color belong to the same community and communities with the same color hue represent clusters of communities with stronger connections. The US has 5 clusters in the following regions: west, north, northeast, southeast, and south. Panels C and D show the degree distribution of the nodes and panels E and F show the edges' length distribution for the communities in panels A and B (see supplementary for detail about the distribution function). Linear behavior of the distributions in log-log axes represents power-law nature of them. Quarantine regulations in the US became effective in late March 2020. Figure 1 shows that quarantine policies did a good job at breaking up some of the connectivity of the areas by reducing the size of the communities, degree of in and out movements to the locations, and the number of large distance movements (by a power-law behavior in range of degree of nodes/length of links). One example, Florida, was connected to the cluster of communities in the northeast of the US during February 23-29. After quarantine, its connections became more local and part of the cluster with neighboring states with closer proximity. See supplementary for more details and community patterns in other weeks.

Applying a community detection algorithm on the network of nodes inside each of the communities reveals the substructure of the communities, exposing the mobility patterns with more detail. In Figure 2, mobility patterns of the US on April 5-11, sub-communities are separated from each other with black lines. Yellow lines represent state borders in upper panel and county borders in bottom panel. Although, community borders in some areas align with administrative borders, in most of areas they deviate significantly from them. This fact reveals that state governments must collaborate with each other, as neighboring states are not necessarily disconnected, based on mobility patterns. A more intelligent contact tracing and quarantine in the suspected areas at the right time can dramatically slow down the acceleration of the pandemic spread to connected areas and reduce its severe impacts. Thus it would be better to carefully define the borders of connected areas based on individual's movements and the scale that policies are applicable and appropriate. To implement contact tracing, it is not enough to trace the places a confirmed case has been, we need to know where the people who were in touch with the infected person were and went.

Unfortunately, due to the delays in applying preventative policies across the US, many areas have seen a large number of cases. By tracking the location of recent active cases and adding them on top of the map of the communities, we can define the risk exposure for the communities. In Figure 3, gray circles show the risk exposure of the communities by counting the number of active COIVD-19 cases in communities and their sub-communities. While doing the analysis in lower resolutions (larger communities) can provide an aggregate view of the world situation, doing the analysis this way will mean the loss of many important details and information. For example, a community may appear be in a bad position when it comes to COVID-19, but when we zoom into the sub-communities, we may see that there is high risk in a few of particular areas, and some areas may have none demonstrating they are safer places and have a better potential to reopen earlier than higher risk areas. The higher the resolution we can provide, the better we can define the local risk levels. Communities with a higher number of confirmed cases need more extensive contact tracing and quarantine policies. Commutes from high to low risk communities can increase the spread of the COVID-19 disease across the society.

By zooming into the map (for example northeast of the US on April 5-11, Figure 4A), interesting facts are observable:

- Areas with no mobility data: There are some urban areas that do not share mobility data, like a Native American community in New York state (21), Figure 5B.
- Isolated communities: Some parts of a community can be geographically disconnected from the rest of the community. This, for example, occurs in university and vacation areas for larger cities.
    - Universities: New York State is the home to many universities. These universities attract people from different areas. The examples of Cornell University and SUNY Cortland, Figure 5C, are two Universities that are located in central New York State yet are isolated sub-communities for New York City. This corresponds with a 2014 investigation (22), which estimated that 65% of all students at Cornell from New York State came from that region of the state.
    - Vacationers: There are vacation spots that individuals from metropolitan regions of one community go to yet are in the middle of different communities. These regions are known for their nice outdoor spaces and somewhat close proximity to the city they are connected to. This phenomenon creates isolated communities in the middle of other communities. Multiple reports have mentioned this pattern occuring (23, 24). In New York City, the Catskill Mountains are one of these escapes, Figure 5D, while for Philadelphia the Poconos serve the same purpose, Figure 5E.
- Sub-communities within other sub-communities: All around the US, there are some areas in which people mostly interact with those immediately around them rather than their nearby urban areas. University campuses are good examples of such sub-communities. As shown in Figure 5F, within the community of upstate/western New York there are specific sub-communities with connections to university campuses. On the left side of the figure, the smaller shape south of Rochester is Rochester Institute of Technology, while the right side of the figure has 2 smaller areas in Syracuse, both associated to Syracuse University. These universities are large and attract many students from the upstate/western New York region.

- Communities that cross state borders: Examples include the area of Philadelphia and southern New Jersey, Figure 5G. The light purple region highlights the Philadelphia community. This community is multi-state and includes parts of northeastern Maryland, northern Delaware, southern New Jersey, and south eastern Pennsylvania. The southern part of the Jersey Shore is a popular travel destination for people from Philadelphia, and the areas in Delaware and Maryland appear to be extensions of the great Philadelphia area.
- Sub-communities in city areas: Racial and income differences, city infrastructure and transportation can be reasons for community formation in city areas. In New York City, these communities are shown by light green in Figure 5H. Brooklyn and Queens have defined sub-communities that are necessary to investigate. Sub-community 1 includes areas of Queens (Long Island City, Astoria, Sunnyside, Woodside, Jackson Heights, Elmhurst, Corona). Sub-community 2 includes parts of northern Brooklyn (Williamsburg, Greenpoint, Maspeth, Middle Village, Rego Park, Forest Hills, Bushwick, Ridgewood, Glendale). Sub-community 3 includes central Brooklyn (Clinton Hill, Bedford-Stuyvesant, Fort Greene, Prospect Heights, Crown Heights, Flatbush, and Canarsie). Sub-community 4 includes areas around Prospect Park (Park Slope, Greenwood Heights, Kensington, Windsor Terrace, Prospect Lefferts Gardens). The public transportation that supports them is different for each area (within Brooklyn division as well). The prospect Park area of Brooklyn is the most wealthy (sub-community 4, smallest subsection). The border between sub-community 4 and 3 on the map can basically be the wealth divide (25). The racial divide between sub-community 3 and 4 can be striking on this map as well (26).

In conclusion, mobility patterns are one of the signs that not only reveal the effectiveness of lockdown policies, but also define areas that are in high risk with regard to the severity of COVID-19 exposure and need for more restriction actions. Mobility networks represent patches that people mostly stay within. This is important because they are also mostly in contact with individuals inside those patches. Lockdown and quarantine policies should attempt to change the mobility patterns and adapt to and strengthen borders of the patches. These policies should eliminate most of the long-distance movements and make them more localized. Patches in the city areas are mostly smaller and in the suburban or rural areas, they become larger. Quantifying movements from and to the patches and restricting commutes between low and high-risk patches can be used to control the spread of coronavirus across various areas and help policymakers and governments to control the pandemic.

**Acknowledgments**
**Funding:** L.H., Y.B-Y, and O.B were supported by the US National Science Foundation (NSF) through the NSF grant 2032536. **Author contributions:** Conceptualization: L.H., Y.B-Y, O.B. Data curation: O.B, D.C. Formal analysis: L.H., O.B., Supervision: L.H., Y.B-Y, O.B. Writing original draft: L.H., A.N., D. C., O. B. Writing, review, and editing: L.H., Y.B-Y. **Competing Interests:** The authors declare no competing interests. **Data source:** SafeGraph. **Data and material availability:** Data are available at: https://www.endcoronavirus.org/mobility-maps.

**Supplementary materials**
Materials and Methods
Supplementary Text
Figs. S1 to S4
References (1-10)

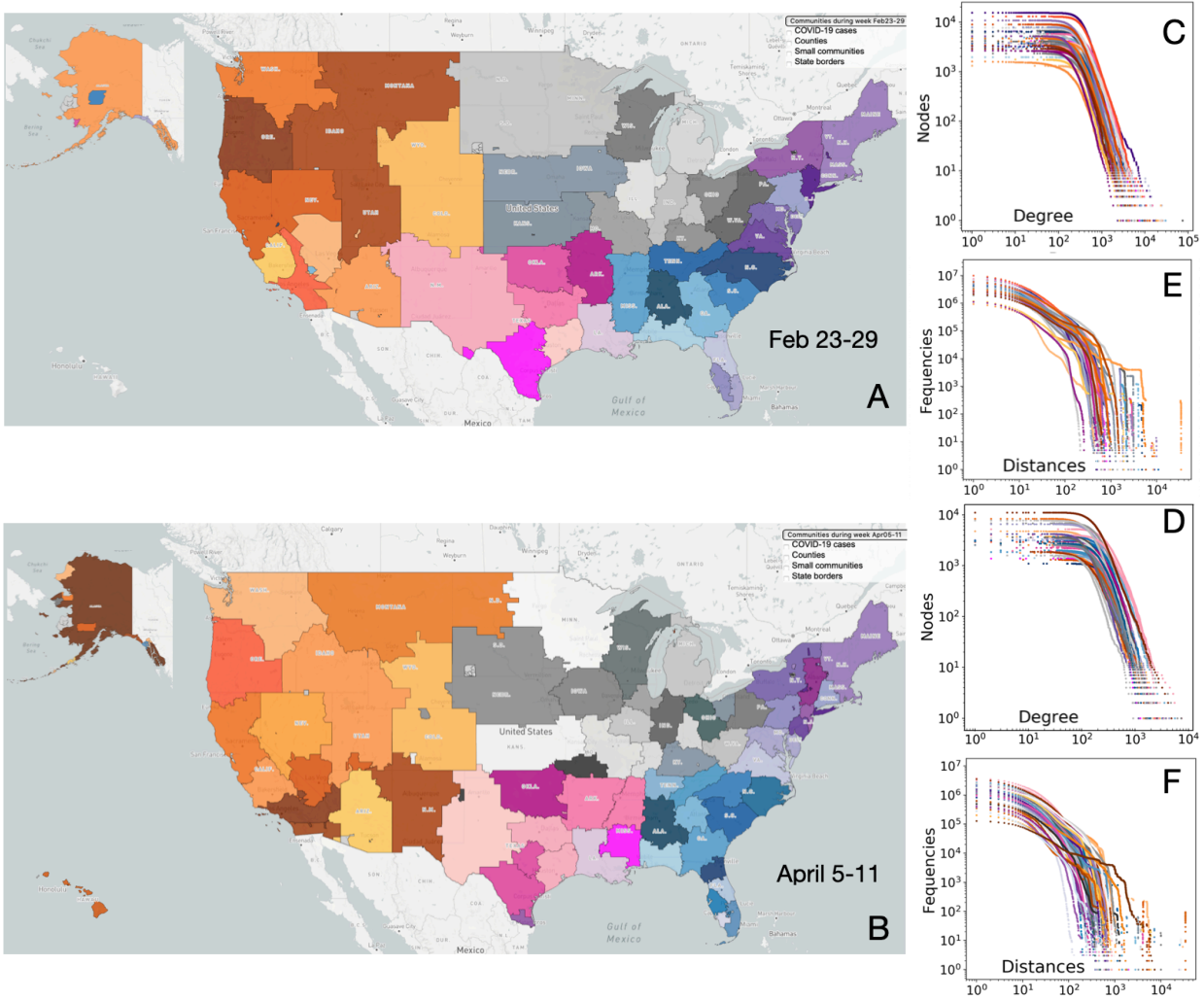

**Fig. 1.** Communities of people that mostly move within the same area on (A) February 23-29 and (B) April 5-11. Communities are shown by different colors. Color hues represent clusters of communities that have higher mobility connections among themselves. (C) and (D) show distribution of in and out links to the nodes inside the communities (degree distribution) of panels (A) and (B). (E) and (F) are the distance distribution of links inside the communities of panels (A) and (B). Axes are logarithmic in the distribution panels indicating that degree of nodes and distances between node s after a threshold decrease by a power low behavior towards larger degrees and distances.

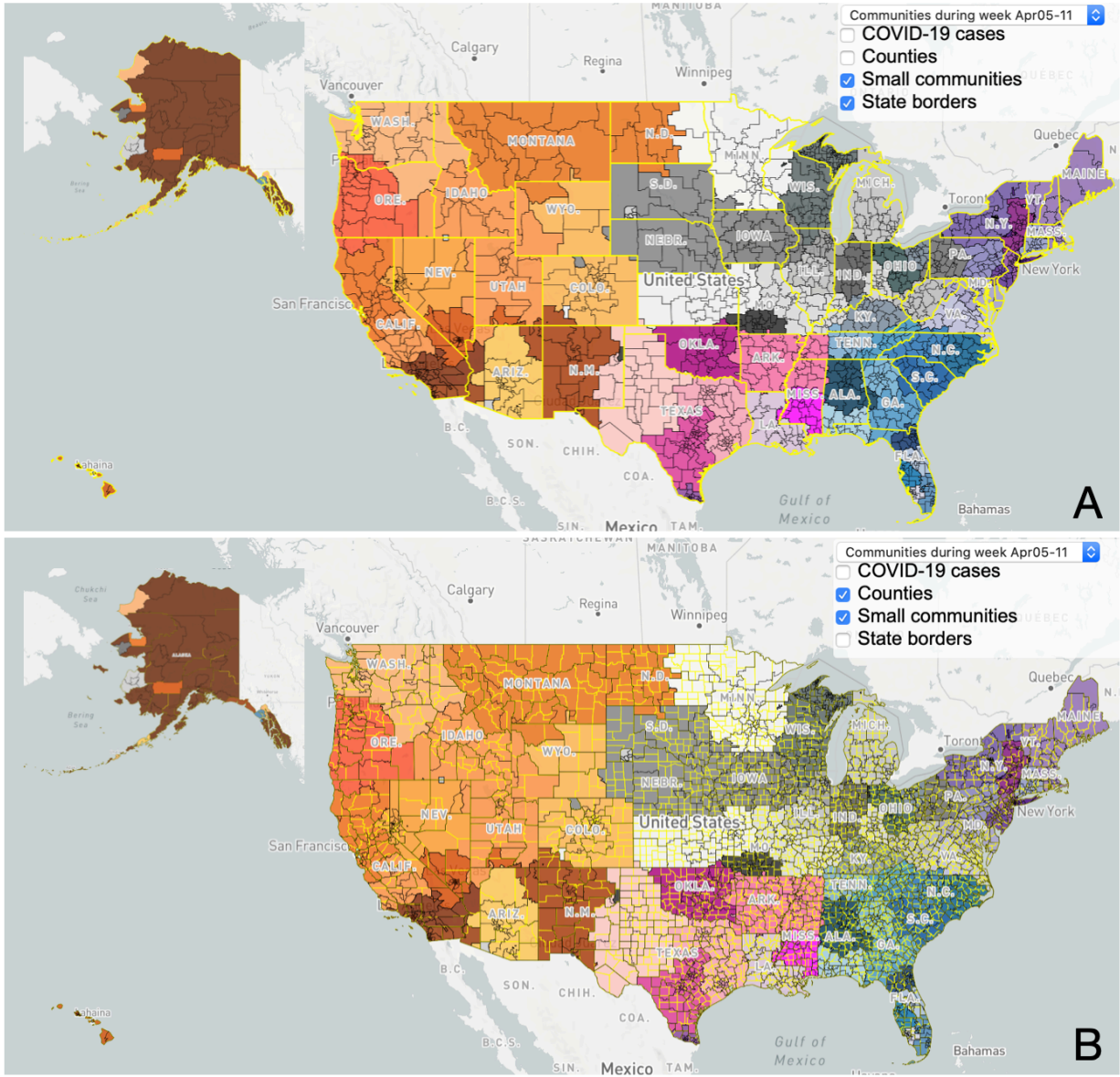

**Fig. 2.** Communities (separated by different colors) and their sub-communities (separated by black lines) of the mobility pattern of the US on April 5-11. Yellow lines show (A) state boundaries and county boundaries (B). Deviation of communities and sub-communities from the administrative borders are clear.

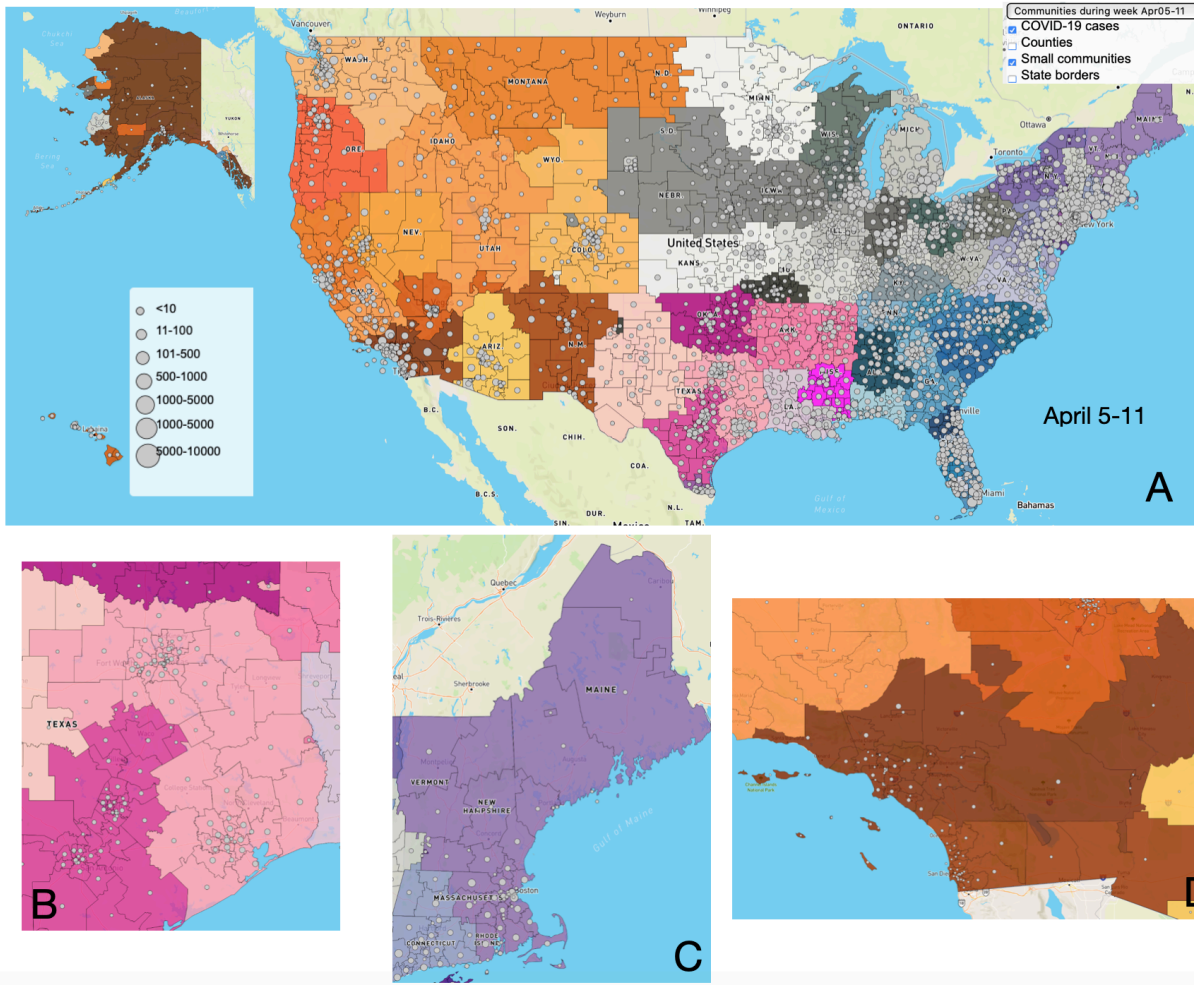

**Fig. 3.** (A) COVID-19 cases on the map of mobility communities of the US on April 5-11, shown by gray circles. (B) and (C) While a community appears to be in bad situation, in higher resolution (sub-communities), some of the areas are in low risk and others in high risk.

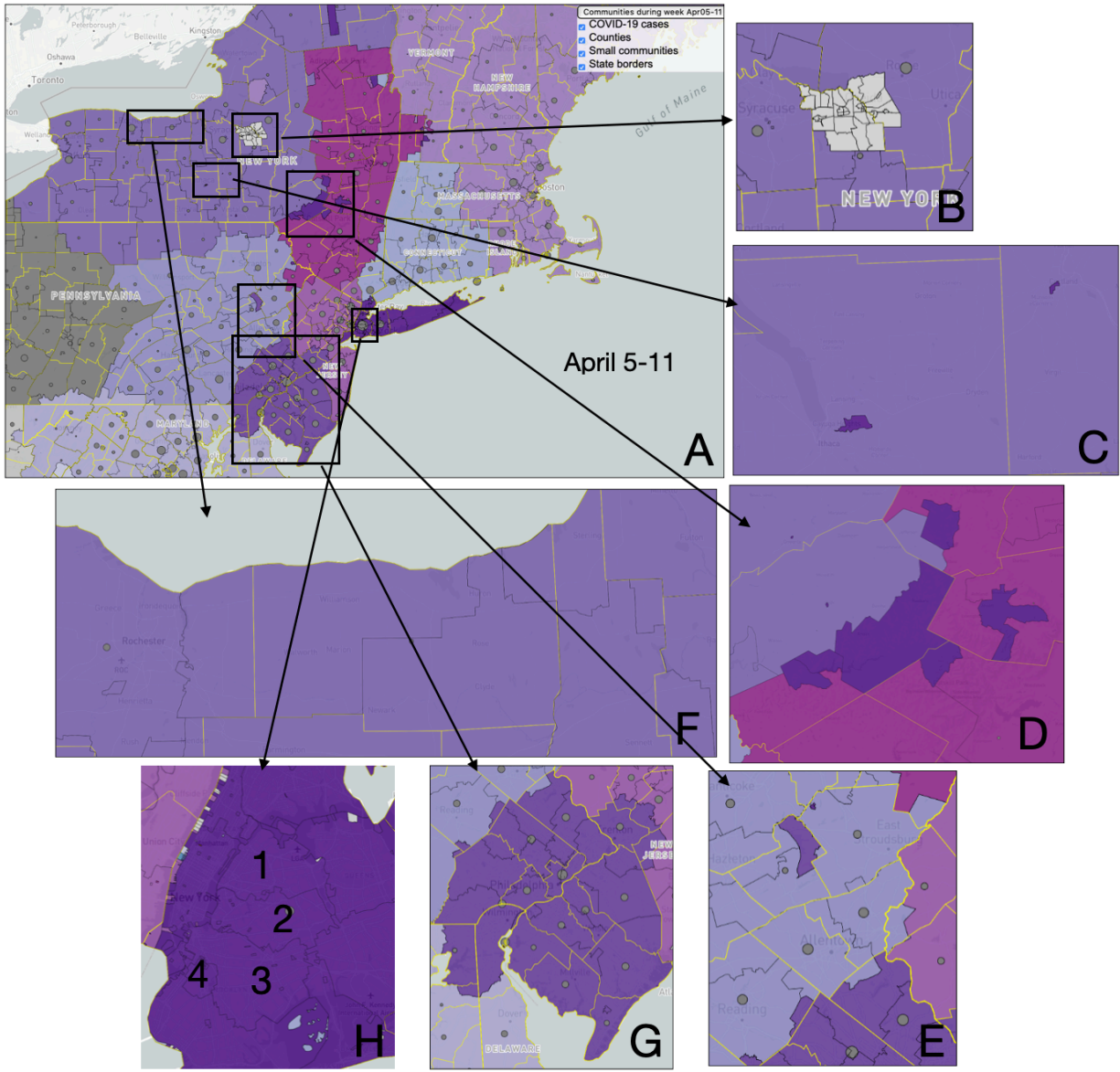

**Fig. 4.** (A) Zoom into the northeast of the US on April 5-11. Special examples of the map: (B) Areas with no mobility data, Isolated communities like (C) universities and (D) and (E) vacation spots, (F) sub-communities within other sub-communities, (G) communities that cross state borders, and (H) sub-communities in city areas.

# Supplementary Materials for

# Strategizing COVID-19 Lockdowns Using Mobility Patterns

Olha Buchel[1], Anton Ninkov[2], Danise Cathel[1],
Yaneer Bar-Yam[1], Leila Hedayatifar[1*]

[1]New England Complex Systems Institute, 277 Broadway St, Cambridge, MA, USA
[2]Faculty of Information and Media Studies, University of Western Ontario, Canada

To whom correspondence should be addressed; E-mail: leila@necsi.edu.

## Materials and Methods
**Data**
    COVID-19 datasets: We use daily confirmed cases time series data from Johns HopkinsUniversity COVID-19 Data Repository. This dataset provides cumulative counts of confirmed cases at county level for the US. By adding the number of active confirmed COVID-19 cases to the map, we define risk exposure for the communities.

    Mobility datasets: We are collecting multiple data sets to extract the mobility networks in the US. In March 2020, technology companies that gather geo-located information on individuals started to share anonymized mobility data to help researchers stop the spread of COVID-19. Each of these datasets covers aspects of an individual's movements. In this work, we used the aggregated mobility datasets by SafeGraph company. SafeGraph provides cellphone data and, for security of the users, anonymizes the data and aggregates them in census block groups (CBG). The original data comes in the CSV files. Each file describes individual CBGs and lists links with weights (number of links) to other census block groups that occurred on a specific day. First, we separate all these relationships and describe them as individual objects. Each relationship has a source, target, date, and weight of interaction. Daily dataframes are combined into weekly dataframes; they are grouped and their relationships are summed. Each census block group in each relationship is augmented with central points derived from census block group polygons.

**Methods**
    Mobility Network: In the mobility network, nodes represent a lattice with cells overlaid on a map of the US. Cells are CBGs used by SafeGraph and are the nodes in the network. Edges represent the movement of an individual from one CBG (node) to another one. Edge's weight represents the number of people who travel between the two CBGs. This network aggregates the heterogeneities of human mobilities in a large-scale representation of social collective behaviors (1).

    Community Detection Algorithm: We analyze social fragmentation by applying the Louvain method (2) with modularity optimization (3) to the mobility network. Communities refer to the regions in which nodes are connected to each other more than the rest of the network. In the Louvain method, in an iterative process, nodes move to the neighboring communities and join them to maximize modularity (M). Modularity is a scalar value $-1 < M < 1$ that quantifies how distant the number of edges inside a community is from those of a random distribution. Values closer to 1 represent better detected communities. Due to the existence of multiple local minima in the Louvain algorithm, some variation in the assignment of nodes may occur between algorithm runs (3,4). To quantify the stability of detected communities and identify areas in which communities overlap with each other, we generate an ensemble of multiple realizations and analyze the borders of patches in all the realizations (5, 6).

    Combining communities into polygons: Each CBG corresponds to a polygon and belongs to a community. We merge and dissolve polygons of CBGs that belong to the same community.

Quantifying community connection: By creating a network that consists of communities as nodes and aggregate mobilities between them as the links, we can define inter-community distances. Using this network structure we can further define clusters of communities with stronger connections. These clusters describe higher level aggregate behavior enabling policy decisions about travel restrictions at the larger scale.

Coloring communities: It is challenging to be consistent in assigning colors to the communities on different weeks, because community numbers assigned by Louvain algorithm are changing. We are able to preserve color hues for mega-communities: purple hues were consistently assigned to northeastern states, blues to southeastern, grays to midwest, pink-magentas to southern communities, and yellow-brown to the western communities.

Degree and Distance distributions: We examined the distribution of the number of nodes versus degree (number of in and out links to the nodes) and number of links versus the length of the links using a survival function. If the degree of nodes/length of links per nodes/links is in range $A = 1,2,3,...,A_{max}$, the survival function counts the frequency of nodes/links that have more than A degrees/lengths, $S(A) = \Sigma_{i=1}^{A_{max}} n_i$. In this work, we plot the distributions for each community separately with colors that match with the color of communities in the shown week, see Figure 1. Notice that the mobility networks are directed and weighted meanings that the links have a direction showing the origin and destination and the links may repeated several times. In general, the shape of the distribution varies depending on inherent traits of the system and may change over time (7, 8).

If communities include a few nodes/links with large degrees/lengths and many nodes/links with a few degrees/lengths, the distribution will be skewed. An extreme skewed distribution is the power-law distribution, in which the frequency of events decreases as a power of size of the events (9,10). Power-law distributions can be described by $N(A) \sim A^{-\alpha}$, in which $\alpha$ quantifies how heavy-tailed the distribution is. On a log-log plot, a power law distribution appears linear, and its slope is equal to $\alpha$. Power-law behavior is also termed "scale-free" because it follows the same relationship at all scales (9).

## Supplementary Text

### Dynamics of mobility patterns

While first signs of the coronavirus pandemic started in late December 2019 from China, and the US government closed its international borders to some more infected countries including China in late January 2020, quarantine regulations in the US at the national scale became effective only in late March 2020. Just after the quarantine, many of large gatherings like conferences and concerts, and places like universities and schools got closed. These actions have changed mobility patterns of individuals drastically. In figures S1 and S2, we show the detected communities from mobility networks during weeks ranging from February to May 2020. While there are not many changes in the communities before and during the first weeks of quarantine (as shown in panels in figure S1) many of large communities were split into smaller communities during April and afterwards (as shown in panels in figure S2).

In figures S1 and S2, clusters of communities are shown with variation in color hues (browns, pinks, grays, blues, purples). It shows that the US has always 5 large clusters of communities. Communities in the west are connected to each other and to Alaska and Hawaii to form a single cluster. In the south, multiple Texas communities along with some neighboring communities are connected to each other. The northeast and southeast of the US form into two large clusters and many of the communities in the north central part of the US are connected to each other. The biggest change after the lockdowns was the disconnection of communities in Florida state from the communities in northeast (purple cluster) during February and early March and from the communities in the west (orange cluster) during late March which were the vacation destination for many people in purple and orange clusters before the lockdowns. While the main location of the clusters have not changed over time and most of the communities in the weeks appear in the same cluster, some communities in the borders of clusters appear in either neighboring cluster.

Most of the communities in the clusters are next to each other. But some of the communities or part of a community are geographically disconnected from the origin cluster. For example, in figure S1, during February and March, there are some small areas in the orange cluster that are connected to communities in the blue cluster before the lockdown, like a blue community in Alaska State and two small blue communities in California State. Analysis shows that many from the blue cluster in the east travel to these areas in the orange cluster in the west. After lockdowns came in place during late March, many of these connections are lost meaning the quarantine policies reduced some long distance travels.

**Movements from and to communities**

While these communities and sub-communities reflect the areas in that individuals mostly move inside the community, there are also many movements from and to the communities from other communities. Movements to a community from high risk communities increase the risk of the exposure to whole the area of the community. So knowing the inter-movements between communities allow us to detect risky long distance movements that can spread the virus across the country. In figures S3, S4 and S5, we quantify connections from and to three example subcommunities to represent differences in mobility patterns in different geographical locations of the US. In all three figures, size of the circles represents the number of movements from other sub-communities to the mentioned sub-community (Upper panels) or from the mentioned sub-community to other sub-communities (bottom panels).

Figure S3 indicates movements from and to a sub-community in the Main State at the northeast of the US (shown with a black arrow) during May 24-30. Total outgoing and incoming links are 21178 and 16935 in that week. While there are almost some movements from and to this sub-community from all sub-communities across the US, most of the movements are from neighboring sub-communities in the northeast.

Figure S4 indicates a sub-community in the New York City (shown with black arrow) during May 24-30. Total outgoing and incoming links are 99574 and 103186 (much larger than the links in sub-community figure S3). According to the figure, there are many movements from that area to the main cities in all the states, specially the ones in the east side of the US. Many movements happened to the sub-communities around the New York City and neighboring states. But the figure also shows a large number of movements to the sub-communities in Florida State. This can be a reason that why Florida experienced a huge outbreak started in late May after the large outbreak in New York City started in late March.

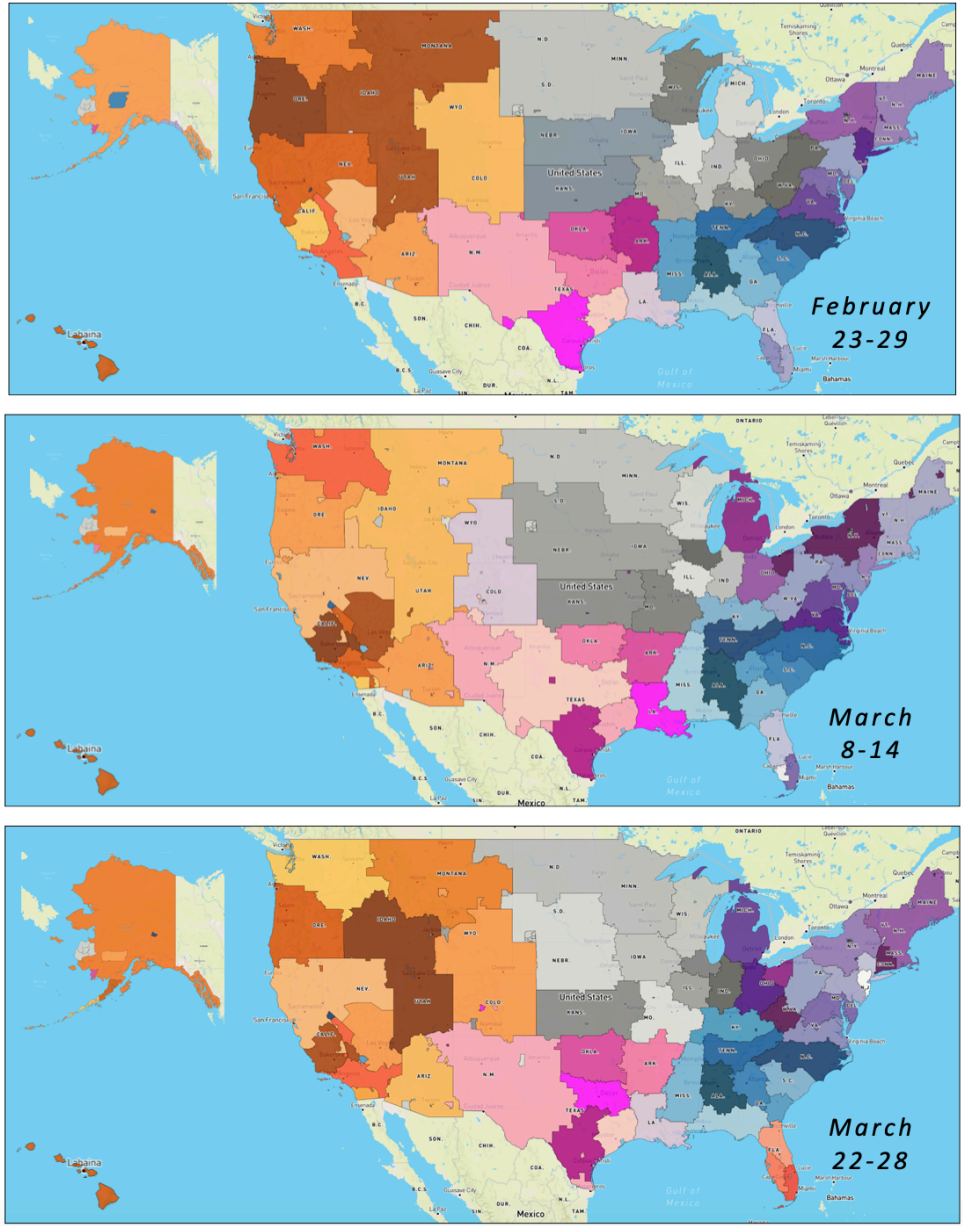

**Fig. S1.** Mobility communities on February 23-29, March 8-14 and March 22-28. Color of communities with the same color hue is chosen randomly.

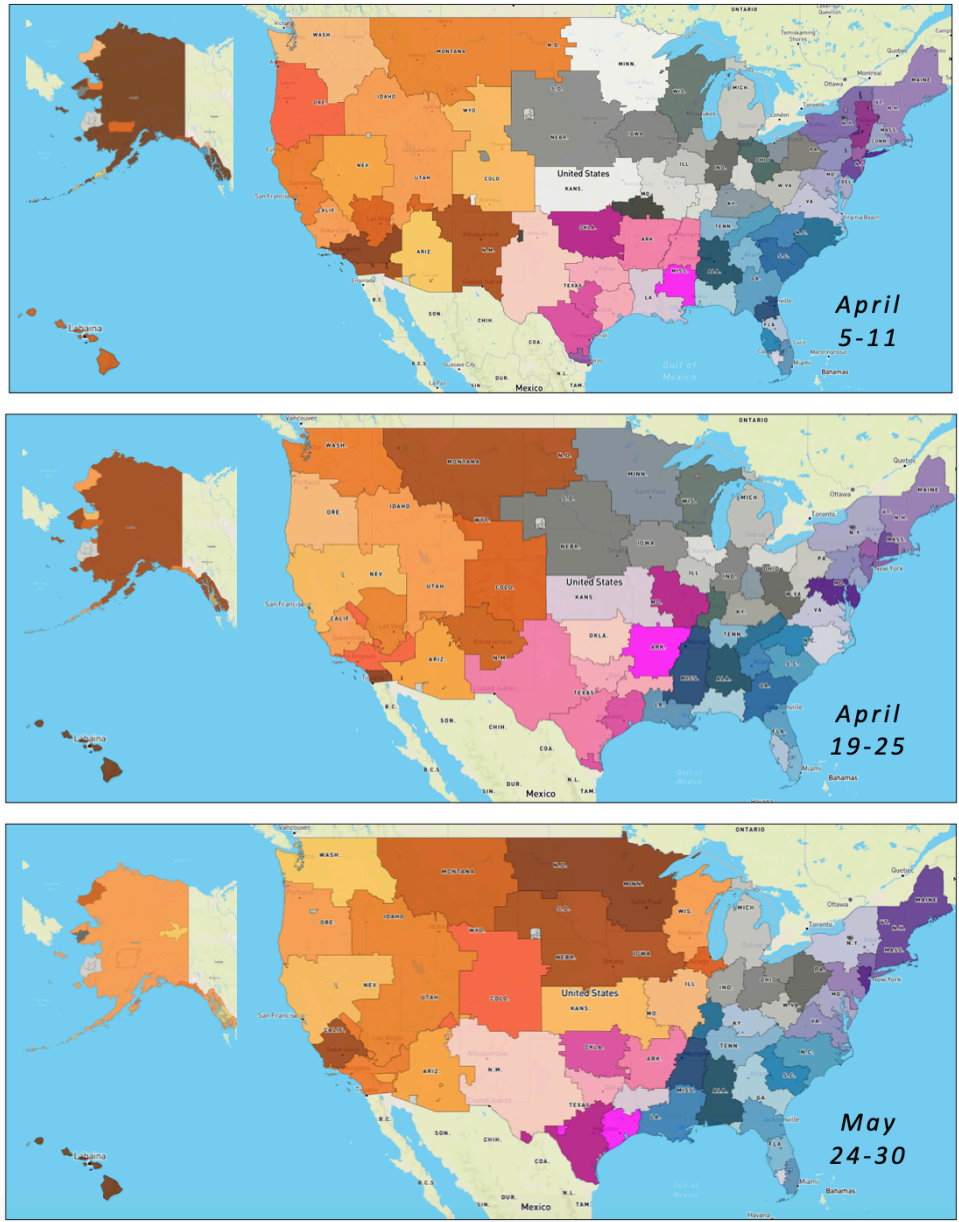

**Fig. S2.** Mobility communities on April 5-11, April 19-25 and May 24-30. Color of communities with the same color hue is chosen randomly.

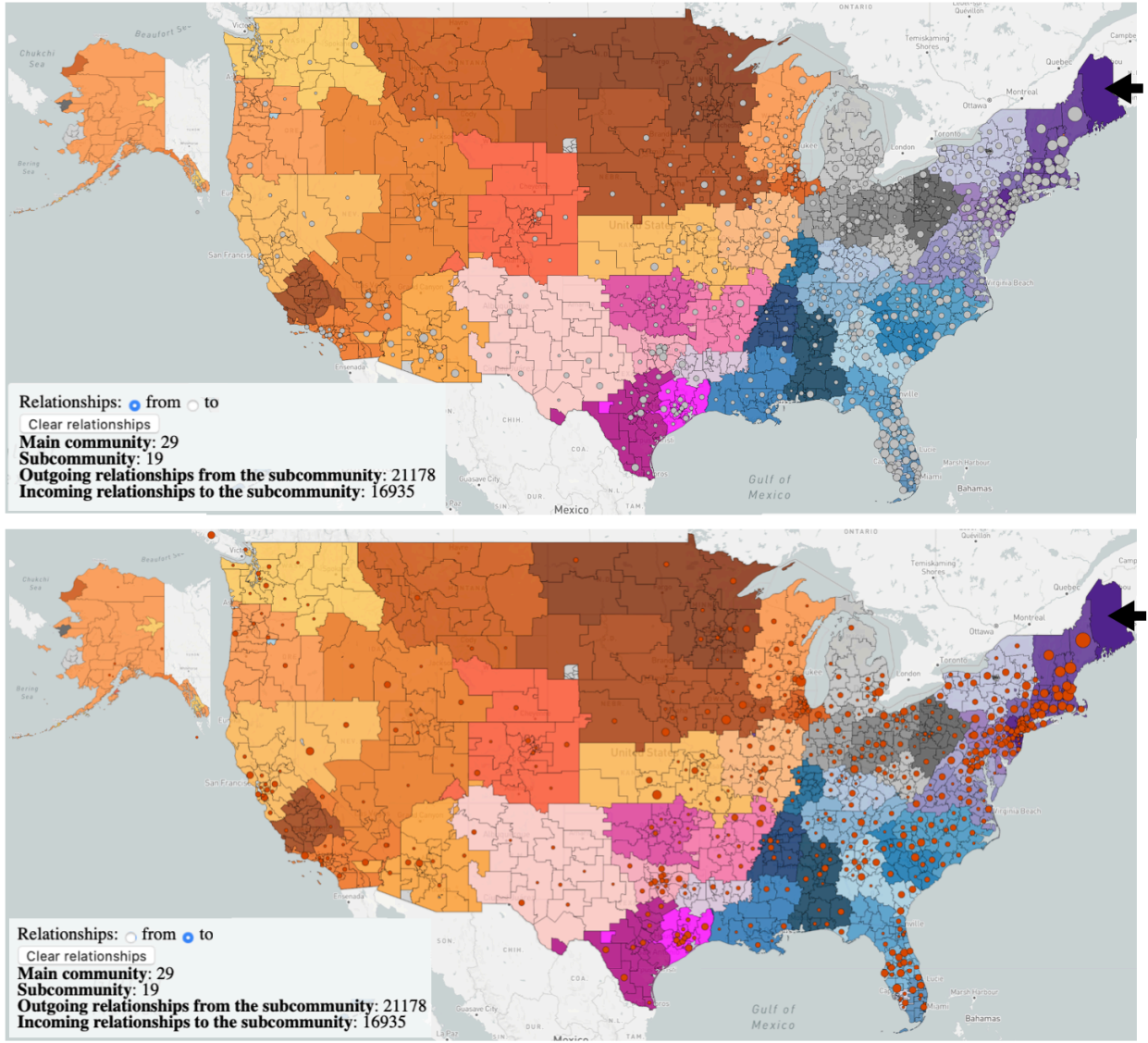

**Fig. S3.** Movements from (upper panel) and to (bottom panel) a sub-community in Main State (shown with an arrow) during May 24-30. Number of outgoing and incoming links to the subcommunity is shown in the legend. Size of the circles shows the amount of movements. This sub-community is located in the northeast of the US and most of the movements from and to it are concentrated in the east the US.

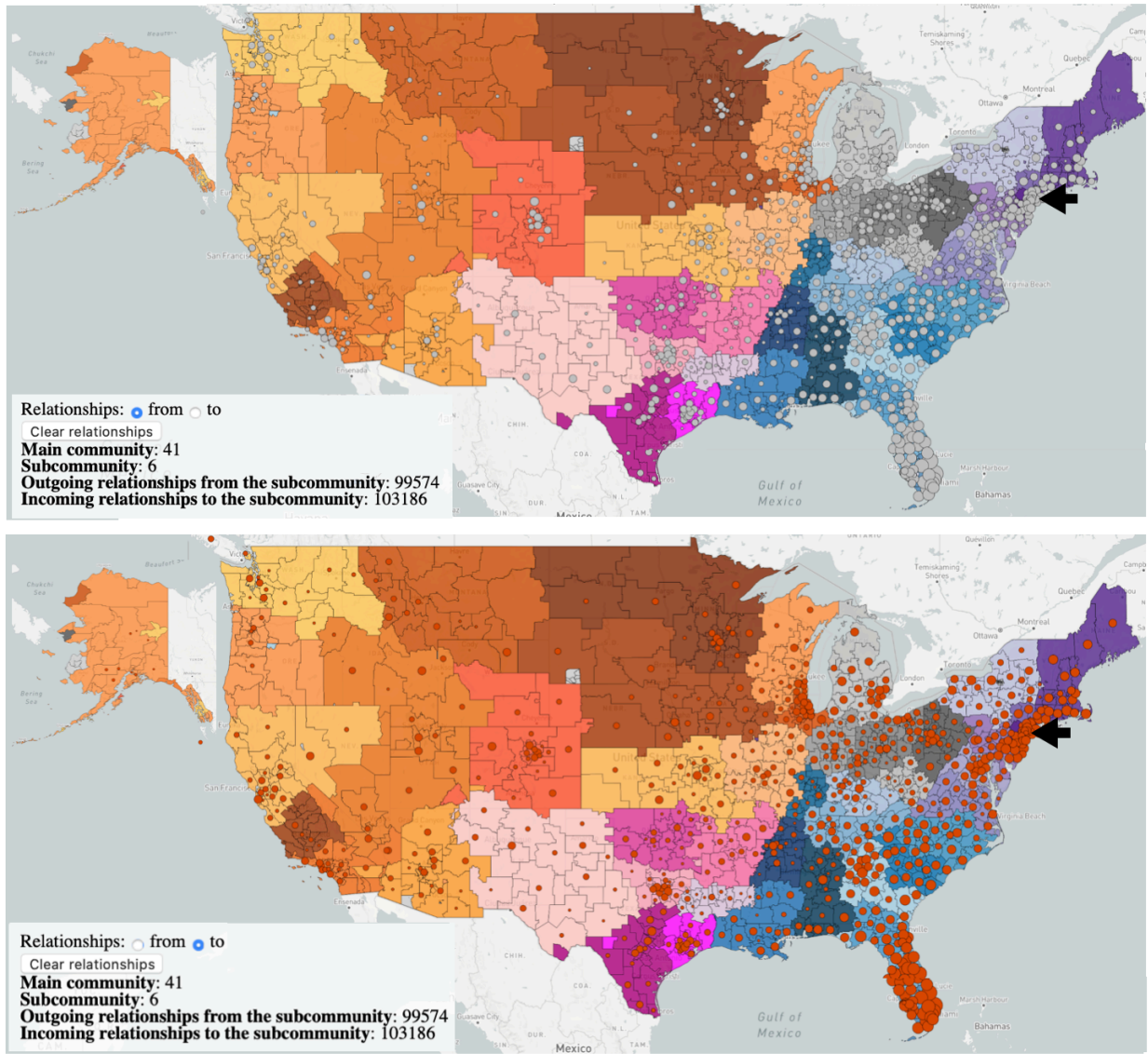

**Fig. S4.** Movements from (upper panel) and to (bottom panel) a sub-community in New York City (shown with an arrow) during May 24-30. Number of outgoing and incoming links to the sub-community is shown in the legend. Size of the circles shows the amount of movements. While many movements have been happened in the east part of the US, most of them are in New York City neighborhood and in Florida state.